\documentclass[letterpaper]{article}
\pdfoutput=1
\usepackage{jheppub}
\usepackage{graphicx}
\usepackage{amsmath}
\usepackage{amssymb}
\usepackage{hyperref}
\usepackage{cleveref}
\usepackage{comment}

\def\GeV{\ \rm{GeV}}
\def\TeV{\ \rm{TeV}}

\newcommand{\msbar}{\overline {\rm MS}}

\title{A Flavorful Factoring of the Strong CP Problem}

\author[a]{Prateek Agrawal}
\affiliation[a]{Department of Physics, Harvard University, Cambridge, MA 02138, USA}
\author[b]{and Kiel Howe}
\affiliation[b]{Fermi National Accelerator Laboratory, Batavia IL 60510,
USA}

\preprint{FERMILAB-PUB-17-562-PPD}

\abstract{
  Motivated by the intimate connection between the strong CP problem and the flavor structure of the Standard Model, we present a flavor model that revives and extends the classic ${m_u=0}$ solution to the strong CP problem. QCD is embedded into a $SU(3)_1\times SU(3)_2 \times SU(3)_3$ gauge group, with each generation of quarks charged under the respective $SU(3)$. The non-zero value of the up-quark Yukawa coupling (along with the strange quark and bottom-quark Yukawas) is generated by  contributions from small instantons at a new scale $M \gg \Lambda_{QCD}$. The Higgsing of $SU(3)^3\to SU(3)_c$ allows dimension-5 operators that generate the Standard Model flavor structure and can be completed in a simple renormalizable theory. The smallness of the third generation mixing angles can naturally emerge in this picture, and is connected to the smallness of threshold corrections to $\bar\theta$. Remarkably, $\bar\theta$ is essentially fixed by the measured quark masses and mixings, and is estimated to be close to the current experimental bound and well within reach of the next generation of neutron and proton EDM experiments. 
} 

\begin{document}
\maketitle

%\newpage
\section{Introduction}

The standard model contains two physical CP violating parameters: (1)
the perturbative CKM phase, 
which originates in the misalignment of the eigenvectors of the Yukawa
matrices $y_u$ and
$y_d$~\cite{PhysRevLett.55.1039},
\begin{align}
  \delta_{CKM} = \arg \det \left[y_u y_u^\dagger , y_d y_d^\dagger \right], 
\end{align}
and (2) the strong CP phase 
\begin{align}
  \bar{\theta} = - \arg \det \left[e^{-i\theta}y_u y_d\right], 
\end{align}
which originates from the combination
of the QCD $\theta$ angle
%topological term $\mathcal{L}\supset \theta \tilde{G}^a_{\mu\nu}G_a^{\mu\nu}$ 
and the determinant of the Yukawas.
Although these two phases appear to be intimately related through
their connection to the Yukawa matrices, $\delta_{CKM}$ is observed to
be $\mathcal{O}(1)$, while current limits give $\bar{\theta}\lesssim
10^{-10}$ \cite{Pospelov:1999ha,Baker:2006ts,Afach:2015sja}. This is
the strong CP problem: how can such a small value of $\bar\theta$ be
explained when the quark sector appears to feel $\mathcal{O}(1)$ CP
violation?
In view of the strong connection of the flavor sector with the strong
CP problem, it is natural to explore its solutions in the context of
models which also generate the flavor structure in the standard
model~\cite{Harnik:2004su,Cheung:2007bu,Calibbi:2016hwq}.
We present such a mechanism in this work. 

One appealing class of solutions to this problem are those that
contain a new anomalous $U(1)_{PQ}$ symmetry.
%When the
%non-perturbative breaking of the PQ symmetry from QCD itself is larger
%by a factor of $\sim10^{10}$ than any other sources of PQ breaking,
%the strong CP problem can be solved.
%
%
The most economical possibility is the ``massless up quark solution'',
where setting  $m_u=0$ at a scale above the QCD scale leads to a
$U(1)_{PQ}$ symmetry.
This is not a priori inconsistent with current algebra
since non-perturbative effects can
generate an effective up-quark
mass~\cite{Georgi:1981be,Choi:1988sy,Kaplan:1986ru,Banks:1994yg} 
(see~\cite{Dine:2014dga} for a review).
In the simplest
extensions of the standard model, non-perturbative QCD effects are only relevant
at the scale $\Lambda_{QCD}\sim\GeV$, and the mechanism can therefore
remove any contributions to $\bar\theta$ generated above the scale
$\Lambda_{QCD}$.  
Unfortunately, the massless up-quark solution is now
strongly disfavored by lattice results, which
find a non-zero $\msbar$ value~\cite{Aoki:2013ldr,Agashe:2014kda},
\begin{align}
m_u
&= \left. 2.3^{+0.7}_{-0.5} \right|_{\mu=2\GeV}
\,.
\end{align}
The significance with which this rules out $m_u=0$ solutions is more
difficult to quantify. Refs.~\cite{Dine:2014dga,Frison:2016rnq} have
recently pointed out some ambiguities and suggested further direct
lattice tests that can support this conclusion.

In this work we consider an extension of the massless up quark solution into models where large non-perturbative effects are generated by embedding QCD as the diagonal subgroup of a $SU(3)^N$ gauge group.  This mechanism for ``factoring" the Strong CP problem was first presented in Ref.~\cite{Agrawal:2017ksf}, where all of the quarks are charged under a single SU(3) factor, and the PQ symmetry is realized by a heavy axion in each sector. In this work, we instead give a flavorful embedding of the quarks in a $SU(3)\times SU(3) \times SU(3)$ gauge group, with each quark generation charged under a separate factor. Each factor contains an independent PQ symmetry implemented by a perturbatively massless quark instead of a heavy axion, and the observed non-vanishing Yukawa couplings are generated entirely by non-perturbative effects at a high scale $M$. These non-perturbative effects can be sizable because although the SM QCD coupling is weak at high scales $M\gg \Lambda_{QCD}$, each individual $SU(3)$ factor can easily be near strong coupling\footnote{We will generalize this model to include
  additional $SU(3)$ factors with no charged matter, making each
  factor more strongly
  coupled, so that the non-perturbative effects can
be made larger still.}.
Higher dimension operators generate
the quark mixing matrix upon the breaking to the diagonal group.
Below the scale $M$ the theory matches
to the standard model with no additional matter. Since in the standard
model $\bar\theta$ is very well sequestered from
$\delta_{CKM}$~\cite{Ellis:1978hq,
Shabalin:1978rs,Khriplovich:1985jr,Dugan:1984qf}, solving the strong
CP problem at the scale $M$ solves it at low energy as long as no new
sources of flavor or CP violation are introduced \cite{Dine:2015jga}.
While $\bar\theta$ is suppressed in this model at tree-level, a
non-vanishing radiative contribution is generated with a size directly
connected to the observed quark masses and CKM angles. Remarkably, the
model predicts $\bar\theta \sim 10^{-10}$, just below the sensitivity
of current EDM experiments and within reach of proposed next
generation neutron EDM \cite{Ito:2007xd,Tsentalovich:2014mfa} and
proton storage ring experiments \cite{Anastassopoulos:2015ura}.

Other models that can explain
$\bar\theta=0$ at tree level in the UV typically require large
discrete symmetries and extensions of the flavor structure, and do not
preserve the radiative sequestering of $\bar\theta$ present in the SM.
For example, in Nelson-Barr models
\cite{Barr:1979as,Nelson:1983zb,Barr:1984qx,BENTO199195}, the
radiative contributions $\Delta\bar\theta$ generally exclude the most
appealing models unless some allowed couplings have unexplained
suppressions or the symmetry structure of the SM is substantially
extended \cite{Dine:2015jga,Vecchi:2014hpa}. 
There are also other mechanisms that introduce new non-perturbative PQ violating
effects at higher energies $M \gg \Lambda_{QCD}$ to solve the strong
CP problem.
Refs.~\cite{Rubakov:1997vp,Berezhiani:2000gh,
Hook:2014cda,Fukuda:2015ana,Dimopoulos:2016lvn}
consider models where the $\bar\theta$ of the SM is related by a $Z_2$
symmetry to a mirror copy of the standard model with
$\bar\theta'=\bar\theta$. Spontaneous $Z_2$ breaking
\cite{Blinov:2016kte} allows the states of the mirror sector to be
decoupled, and non-perturbative mirror $SU(3)'$ effects to become
strong at a scale $\Lambda_{QCD}' \gg \Lambda_{QCD}$ and
simultaneously relax $\bar\theta'$ and $\theta$ either with a
heavy-axion
\cite{Rubakov:1997vp,Berezhiani:2000gh,Hook:2014cda,Fukuda:2015ana,Dimopoulos:2016lvn}
or a heavy perturbatively massless quark \cite{Hook:2014cda}. These
theories are significantly constrained by the cosmology of the mirror
sector and new colored TeV-scale particles.
Another possibility is that the SM QCD itself becomes embedded in a
strongly coupled gauge group at high
energies--Refs.~\cite{Holdom:1982ex, Holdom:1985vx, Dine:1986bg,
Flynn:1987rs, Choi:1998ep} considered the possibility that extra
matter causes QCD to run back to strong coupling at a scale $M$ where
it is embedded in a larger gauge group, e.g. $SU(3+N)$. In general to
obtain sizable effects these models also require the addition of new
dynamics breaking the chiral symmetries, and contain new CP violating
phases which cause a misalignment between the non-perturbative
violations of the PQ symmetry at  $\Lambda_{QCD'}$ and
$\Lambda_{QCD}$, spoiling the solution to the strong CP problem
\cite{Dine:1986bg}.

\section{Massless Quark Solution in QCD: the baby version}
\label{sec:two-flavor}

We start from a simpler version of the standard model with only a
single generation of quarks -- the $SU(2)$ doublet $q=(u,d)$ 
and two singlets $u^c, d^c$ -- charged as in the
standard model.
We include an $SU(2)$ doublet Higgs $H$  and assume a UV cut-off $\Lambda_{UV}$. 
We make use of an
anomalous $U(1)_{PQ}$ symmetry under which only $u^c$ transforms, 
\begin{align}
u^c \rightarrow e^{i\alpha} u^c
\end{align}
which forbids an up Yukawa coupling at the
perturbative
level (more precisely, we assume that the dominant source of PQ breaking is from non-perturbative effects within the effective theory far below the scale $\Lambda_{UV}$). The relevant terms in the Lagrangian are
\begin{align}
\mathcal{L}_{\Lambda_{UV}}
&\supset 
\frac{\alpha_s }{8\pi} \theta\,\tilde{G}_{\mu\nu}^aG^{a,\mu\nu} 
+ y_d q H^\dagger  d^c
\end{align}
The $U(1)_{PQ}$ symmetry and a chiral rotation of $d^c$ can be used
to remove the topological phase $\theta$ and the phase of the
non-vanishing Yukawa coupling $y_d$. Therefore there is no physical CP
violating parameter, $\bar\theta=0$. This is effectively
the massless
up quark solution to the strong CP problem.

Non-perturbative $SU(3)$ effects violate the anomalous $U(1)_{PQ}$, so
non-perturbative effects suppressed as $\sim e^{-2\pi/\alpha_s}$  will
generate a non-vanishing effective $y_u$ coupling at energies below
$\Lambda_{UV}$. In the weak coupling limit, the dilute instanton gas
approximation captures the leading non-perturbative effects, and the
instantons can be integrated out to generate an effective Lagrangian
for the fermions \cite{tHooft:1976snw,Andrei:1978xg,Choi:1988sy}. For
$SU(3)$ with two flavors of quarks, both four-fermion and bilinear
terms are generated from single-instanton effects,
\begin{align}
  \mathcal{L}_{inst} 
  &= 
  \int_{\rho=\Lambda_{UV}^{-1}}^{\rho=\Lambda_{IR}^{-1}} 
  \frac{d\rho}{\rho} e^{i\theta} 
  D\left[\alpha_s(1/\rho)\right]
  \left.\left(
  -c_0 y_d^*  q H u^c 
  + c_0 (2\pi^2)  \rho^2 
  (u^\alpha d^c_\alpha d^\beta u^c_\beta
  - d^\alpha d^c_\alpha u^\alpha u^c_\alpha)
  \right)
  \right|_{\mu=\rho^{-1}}
  \label{eq:Linst2}
\end{align}
where $\alpha,\beta$ are QCD indices and the dimensionless instanton
density is
\begin{align}
D[\alpha]
  &=
  D_0 \left(\frac{2\pi}{\alpha}\right)^6
  e^{-\frac{2\pi}{\alpha}}
\end{align}
which features the
non-perturbative exponential suppression factor at weak coupling. The
analytic constants are $D_0 \approx
0.02$ and $c_0\approx 1.79$~\cite{Choi:1988sy}. The couplings in the integrand are evaluated at
the scale $\rho^{-1}$  (higher order corrections can be found in
Ref.~\cite{Dine:2014dga}). Higher dimension operators are suppressed
by further powers of $D[\alpha]$ , and  $D[\alpha]\sim 1$ signals the
breakdown of the dilute instanton gas approximation. 

The effect of instantons on the Yukawa couplings can be conveniently
described as a non-perturbative contribution to the running of the
Yukawa couplings \cite{Choi:1988sy},
\begin{align}
  \frac{d}{d \ln \mu}
  \begin{pmatrix}
    y_u & 0 \\ 
    0 & y_d 
  \end{pmatrix} 
  &=
  -c_0 D[\alpha(\mu)] e^{i\theta}
  \begin{pmatrix} 
    y_d^* & 0 \\ 
    0 & y_u^*
  \end{pmatrix} 
  \label{eq:yrun}
\end{align}
Recall that the perturbative contributions to the
running of Yukawas are multiplicative, and are negligible here.
Now that non-perturbative effects are included,  $y_u\neq 0$ is
generated and the PQ-symmetry appears to be violated perturbatively in
the low energy effective Lagrangian. However,  the physical CP angle
$\bar\theta$ remains vanishing: the non-perturbatively generated $y_u$
has just the right phase to allow the $\theta$ angle and the phase in
$y_d$ to be simultaneously rotated away, as is clear from
\cref{eq:Linst2}. 

\begin{figure}
\centering
\includegraphics[width=0.65\textwidth]{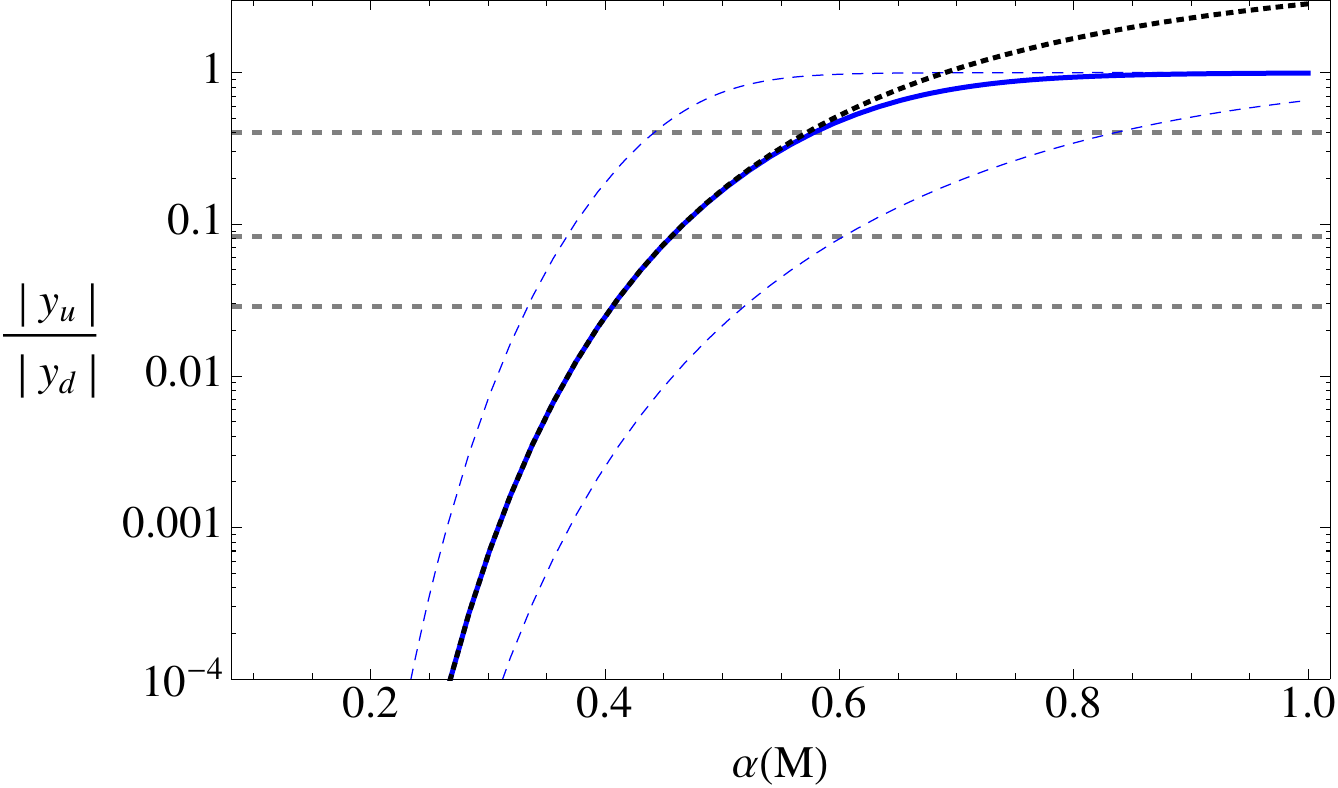}
\caption{The ratio of Yukawa couplings generated by instanton effects
  as a function of $\alpha=g^2/4\pi$ evaluated at the scale
  $M$ in the 2-flavor model assuming a vanishing
  perturbative value for one of the couplings. The solid curve cuts
  uses the full solution to \cref{eq:yrun}, cutting off the running at
  $\mu=M$. The dashed curves cut off the instanton
  effects at $\mu = M/2, 2M$, and are shown
  as a rough estimate of the theoretical uncertainty. The dotted line
  is the single-instanton approximation \cref{eq:singleinstanton}.
The horizontal dashed lines are, from top to bottom, the experimental
values of $y_u/y_d$, $y_s/y_c$, and $y_b/y_t$.
}
\label{fig:yuydRatio}
\end{figure}

Two-flavor QCD is asymptotically free and the instanton density grows
in the IR. If $SU(3)$ is Higgsed at the scale $M$, the
instanton contribution to $y_u$ is cut-off and dominated by instantons
of size comparable to the Higgsing, $\rho^{-1}\sim M$ (we will discuss
the nature of the Higgsing sector in the following section). Using the
one-loop running of the gauge coupling $d\alpha^{-1}=\frac{b}{4\pi} d
\ln{\mu}$, with $b=29/3$ for 2-flavor QCD, the  linear solution to the
running \cref{eq:yrun} gives
\begin{align}
  \frac{|y_u|}{|y_d|}
  &=-\frac{2 c_0 D_0}{b}\int_{2\pi /\alpha(\Lambda_{UV})}^{2\pi/ \alpha(M)}  
  \left(\frac{2\pi}{\alpha}\right)^6 
  e^{-\frac{2\pi}{\alpha}} 
  d(2\pi/ \alpha) \approx \frac{2 c_0 D_0}{b}\Gamma(7, 2\pi \alpha(M))
  \label{eq:singleinstanton}
\end{align}
where we have assumed $\alpha^{-1}(\Lambda_{UV}) \ll 1$ for the last
equality, and $\Gamma(n,x)$ is the upper incomplete $\Gamma$-function.
\Cref{fig:yuydRatio} shows the ratio $|y_u|/|y_d|$ after
integrating out effects above $M$ as a function of the QCD coupling at
the scale of Higgsing, $\alpha(M)$.  As $\frac{|y_u|}{|y_d|}$ approaches $\sim 1$,
multiple-instanton effects captured by higher order solutions to
\cref{eq:yrun} become important, and the ratio asymptotes to $|y_u|/|y_d|=1$.  For $\alpha(M)
\sim 0.4-0.8$ an $\mathcal{O}(1)$ ratio can be generated as required by the observed light quark masses. In this regime the dilute instanton gas approximation is only a
qualitative picture of the non-perturbative QCD effects, but strongly
suggests that they are $\mathcal{O}(1)$ and that a viable ratio $|y_u|/|y_d|$ can be realized before the theory enters the chiral-symmetry breaking phase which would be expected to occur at $\alpha(M) \gtrsim 0.7-1$ \cite{Roberts:1994dr,Appelquist:1997gq}.  As the theory flows to weak coupling at scales above M, the PQ violating effects are rapidly suppressed. For example for
$\alpha \approx 0.1$, as in the SM near the weak scale, the
non-perturbative  contribution to $y_u$ is  $|y_u|/|y_d|\lesssim
10^{-16}$.

This simple 2-flavor example shows that instanton effects can generate
large non-perturbative contributions to a perturbatively vanishing
Yukawa coupling. In fact such effects are known to be important near the
scale of QCD confinement, $\Lambda_{QCD}$,
in the standard model, as reviewed in~\cite{Dine:2014dga}.
However,  as mentioned
above, lattice results strongly disfavor a massless up quark solution
to the strong CP problem in the SM.

The suppression of this effect in the SM is partly due to the fact
that the strange quark is also relevant at $\Lambda_{QCD}$, and
instanton contributions to $m_u$ are further suppressed by $m_s$. In
fact, 2+1 flavor lattice QCD results fully include all instanton
configurations and can be interpreted as a  calculation of the 2nd
order term in the Chiral Lagrangian giving an effective up-quark mass
proportional to $m_d^* m_s^* / \Lambda_{QCD}$ -- these results suggest
that the size of the desired non-perturbative effect is only $\sim
10$--$40\%$ of the experimentally required value \cite{Dine:2014dga}.

So, although qualitatively non-perturbative effects in the SM near the
scale $\Lambda_{QCD}$ are nearly the right size to allow $m_u=0$
solution to the strong CP problem, quantitatively the possibility is
strongly disfavored by precision lattice results.  In the following
section we will describe an extension to the SM in which
non-perturbative effects can become important again at a high energy
scale $M\gg\Lambda_{QCD}$, and these additional contributions allow a
solution to the strong CP problem reminiscent of the massless up quark
solution. 

\section{Massless Quark Solution in QCD: the real thing}
\label{sec:ThreeGen}
\begin{figure}
\centering
\includegraphics[width=2in]{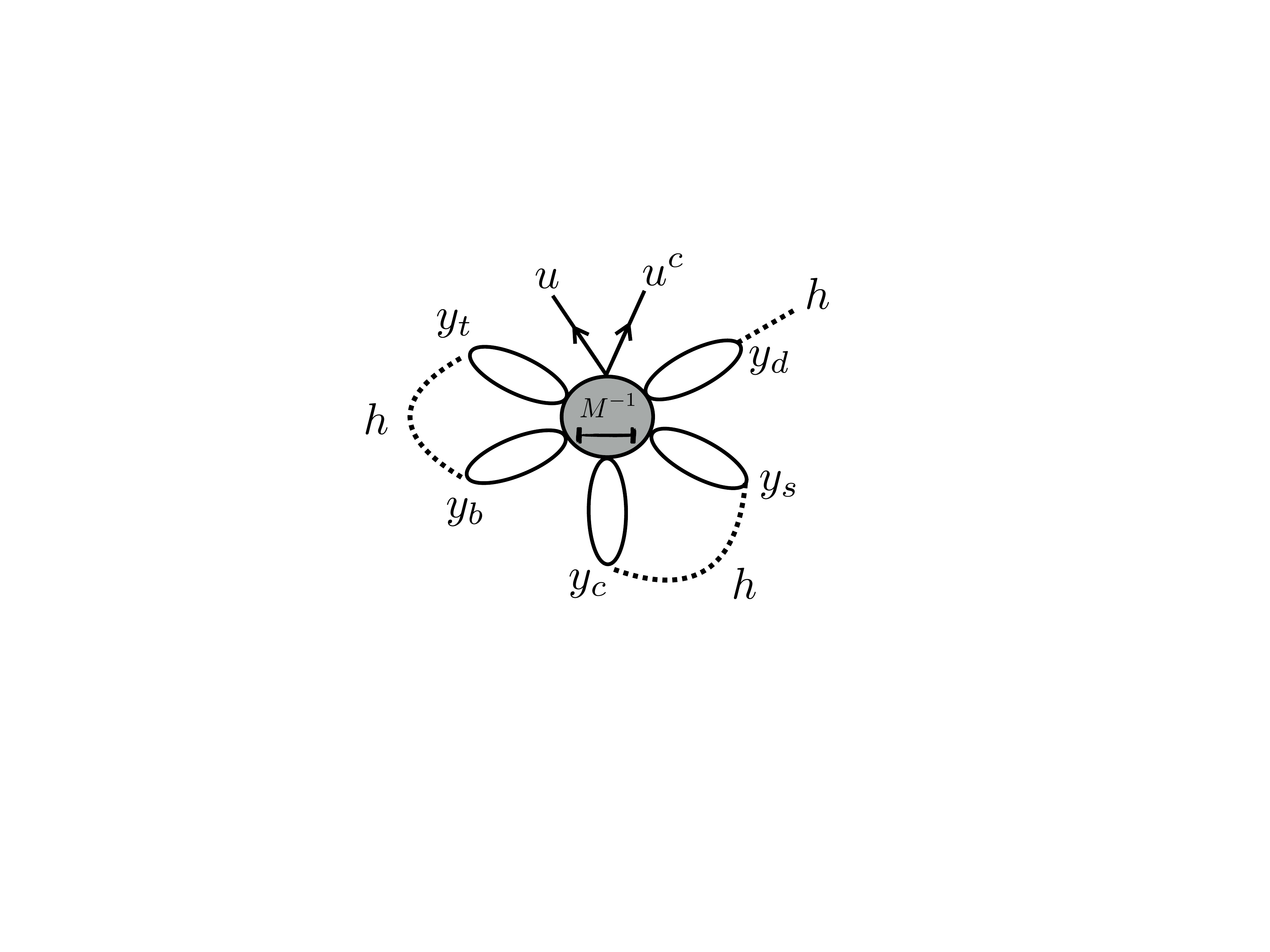}
\caption{An instanton contribution to the up quark mass in the SM at high energies. All six quark flavors appear, and the non-vanishing contributions are proportional to the products of all the Yukawa couplings. For $M \gg v$, diagrams with the Higgs looped off are more important than Higgs vev insertions. }
\label{fig:SixFlavorInstanton}
\end{figure}

Going beyond the illustrative two-flavor example, there are two
challenges to generating a large non-perturbative contribution to the
Yukawa couplings at a new scale $M \gg \Lambda_{QCD}$. The first is
that QCD must be embedded in a strongly coupled theory at the scale
$M$ so that non-perturbative effects are important, but must match to
the weak coupling of QCD in the standard model at high energies, e.g. $\alpha_s(1000 \TeV) \approx 0.05$. The second challenge is that at high energies in QCD, all
three generations of quarks are relevant, leading to further Yukawa
suppressions of high energy contributions from instantons at small
sizes $\rho^{-1} \gg v$. For example, as illustrated in
\cref{fig:SixFlavorInstanton}, the high energy contributions to $y_u$
in the 3-generation SM are further suppressed as 
\begin{align}
|y_u|/|y_d| \sim
\frac{|y_c y_s y_t y_b|}{(16\pi^2)^2}
\end{align}
because the explicit breaking
of each non-anomalous $U(1)_{PQ}$ by the Yukawa couplings must be felt
to generate $y_u \neq 0$.

\begin{figure}
\centering
\includegraphics[width=3in]{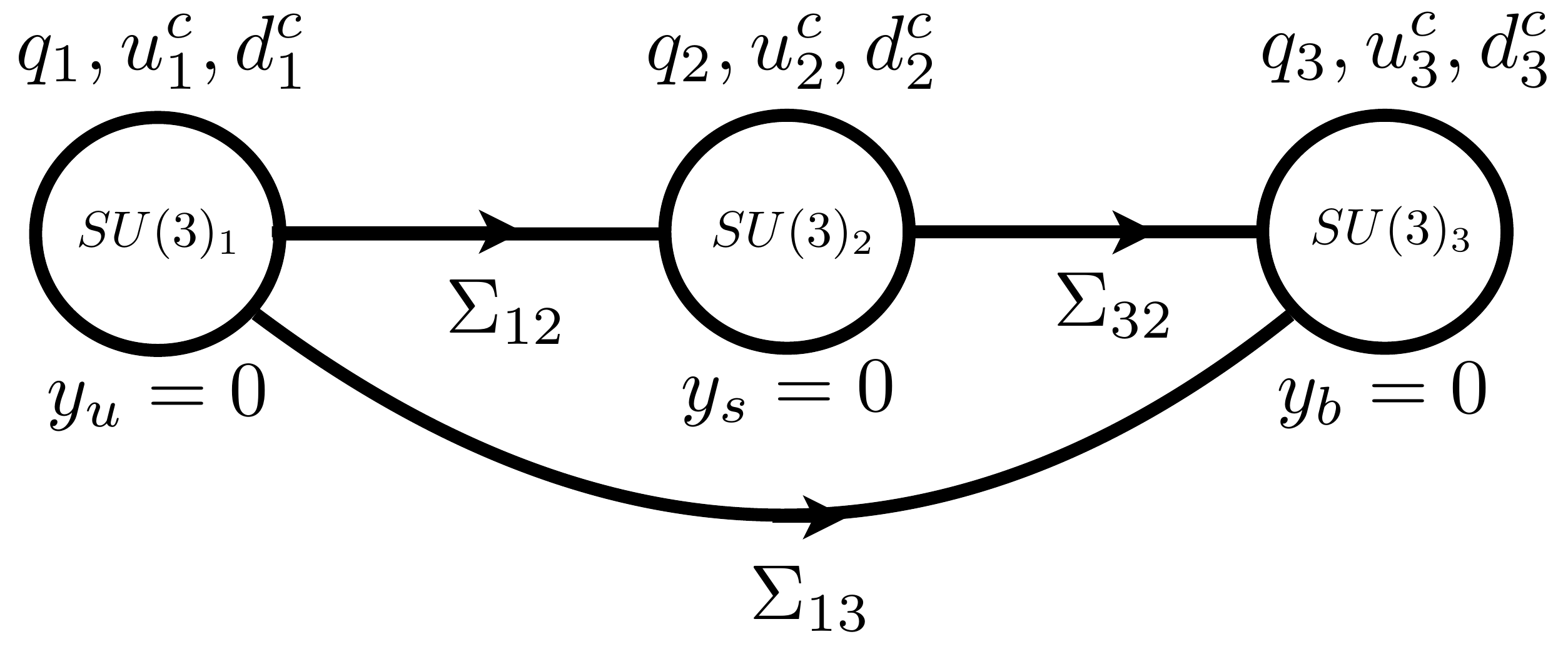}
\caption{The model is a 3-site $SU(3)_1 \times SU(3)_2 \times SU(3)_3$ theory, with one generation charged under each $SU(3)$ factor. The link field $\Sigma$ vevs break the gauge group down to the diagonal  SU(3) of the standard model. One quark in each generation obtains its mass from non-perturbative effects, making the $\bar\theta$ angle in each individual gauge factor unphysical. }
\label{fig:threesite}
\end{figure}

Both these challenges can be solved by embedding the standard model
$SU(3)_c$ into a $SU(3)_1 \times SU(3)_2 \times SU(3)_3$ product gauge
group above the scale $M$, as depicted in \cref{fig:threesite}. Each
generation of quarks is charged under a separate $SU(3)$ factor. The
theory will be Higgsed at the scale $M$ to the diagonal gauge group by
bifundamental scalar fields, as discussed in more detail in the following section. The unbroken diagonal $SU(3)_c$ group's coupling is 
\begin{align}
  \frac{1}{\alpha_s}
&= 
\frac{1}{\alpha_{s_1}}
+\frac{1}{\alpha_{s_2}}
+\frac{1}{\alpha_{s_3}}
\label{eq:weakcoupling}
\,,
\end{align}
allowing  to match to the weakly coupled SM QCD even when each
individual factor is more strongly coupled. 

\begin{figure}
\centering
\includegraphics[width=4in]{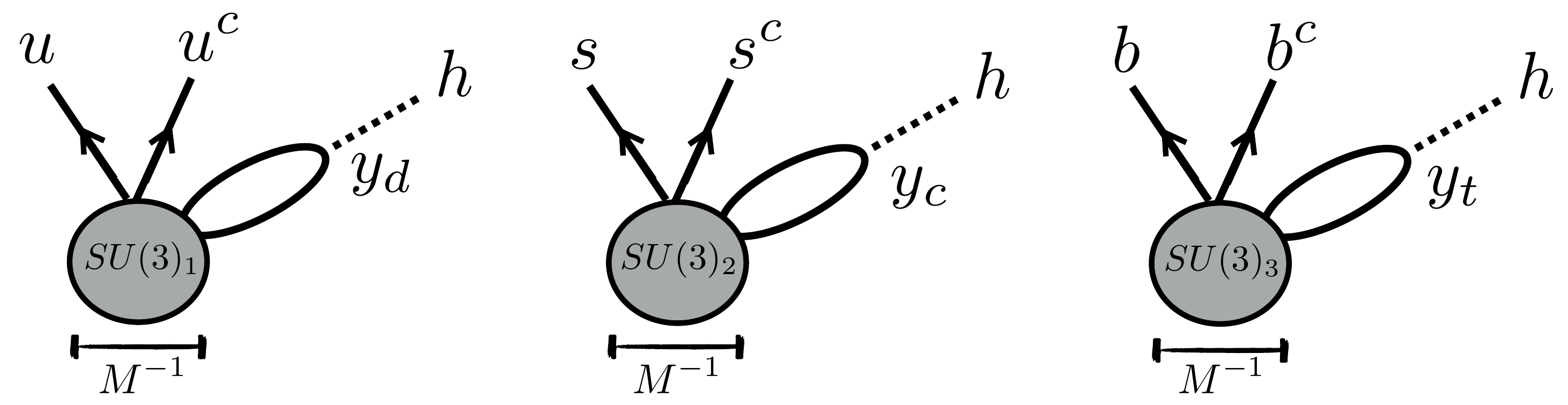}
\caption{Schematically, high energy instantons act separately at scales above $M$ to generate one of the Yukawa couplings in each individual $SU(3)$ factor. }
\label{fig:manyFlavorInstanton}
\end{figure}

Since there are now three separate $SU(3)$ factors, there are now
three separate $\theta$ problems! Fortunately, all the $\theta$ angles
can be made unphysical if there is an independent anomalous
$U(1)_{PQ}$ symmetry in each sector.  The minimal realization of this
PQ symmetry involves a perturbatively massless quark in \emph{each}
sector. Since the non-perturbatively generated Yukawa couping is
always smaller than the unprotected Yukawa, a natural choice is to
choose PQ symmetries that enforce $y_u=0, y_s=0, y_b=0.$ Above the
scale $M$ of Higgsing, each site behaves as the two-flavor model of
~\cref{sec:two-flavor}. Schematically, the generation of the Yukawa
couplings is depicted in~\cref{fig:manyFlavorInstanton}.
From~\cref{fig:yuydRatio}, we can read off the size of the gauge
couplings at the scale $M$ that are necessary for the instantons in
each factor to generate the observed Yukawa ratios:
\begin{eqnarray}
y_u / y_d \approx 2/5  
& \rightarrow & \alpha_{s1}(M) \approx 0.45-0.85  
\nonumber \\ 
y_s / y_c \approx 1/12 
& \rightarrow & \alpha_{s2}(M) \approx 0.36-0.6  
\nonumber \\
y_b / y_t \approx 1/35 
& \rightarrow & \alpha_{s3}(M) \approx 0.33-0.55 
%\nonumber \\
\end{eqnarray}
\Cref{eq:weakcoupling} then gives the coupling of the unbroken
diagonal group at the matching scale  $\alpha_s(M) = 0.12-0.22$.
Flavor constraints will require us to match to the SM at a scale
$M\gtrsim 1000\TeV$ where $\alpha_s(1000\TeV) = 0.05$, so it appears
unlikely that this minimum ${SU(3)_1\times SU(3)_2 \times SU(3)_3}$
model is viable unless our dilute instanton calculation significantly
underestimates the size of non-perturbative effects. 

\begin{figure}
\centering \includegraphics[width=4in]{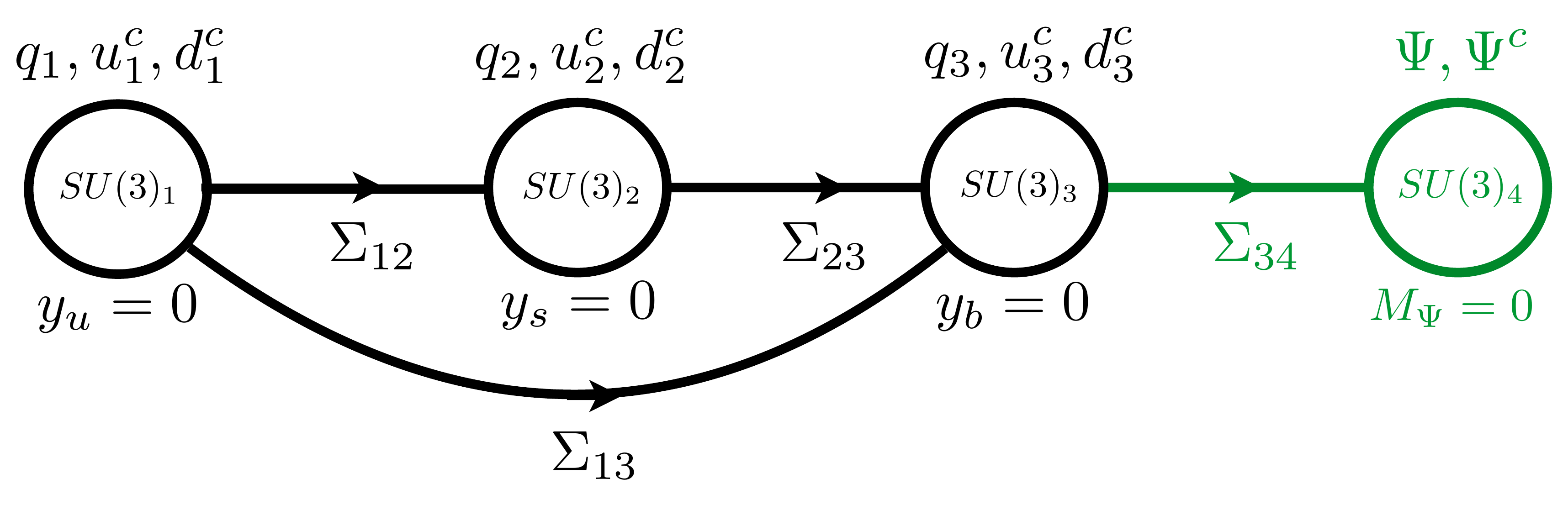} 
\caption{The
  3-site $SU(3)_1 \times SU(3)_2 \times SU(3)_3$ theory of
  \cref{fig:threesite} extended to contain an extra site with a
  more weakly coupled $SU(3)_4$ factor. There is no chiral matter at
  this site, and the $ \theta_4$ angle is removed by an anomalous
  $U(1)_{PQ}$ symmetry of a single vector-like quark species
$\Psi,\Psi^c$.  While $M_\Psi=0$ perturbatively, non-perturbative
effects violating the PQ symmetry generate $M_\Psi \neq 0$. }
\label{fig:foursite}
\end{figure}

One way to overcome this obstacle is to enlarge the product gauge
group to $SU(3)_1 \times SU(3)_2 \times SU(3)_3 \times SU(3)_X^N$,
where the extra gauge factors do not contain chiral matter and
therefore can remain more weakly coupled.  Removing the $\theta$ angle
in these extra factors will involve introducing a PQ symmetry at each
new site, as shown for example in~\cref{fig:foursite}. For
example, the $\theta$ parameter in the extra sites can be removed with
very heavy axion degrees of freedom as discussed in
Ref.~\cite{Agrawal:2017ksf}. Another simple viable possibility is to
add $N=3$ or $4$ sites, each with a colored vectorlike particle
$\Psi_X, \Psi_X^c$ and $M_\Psi =0$ perturbatively to realize a PQ
symmetry. Instantons in each factor generate a mass $M_\Psi \sim
D(\alpha(M)) M$. For $N=4$ and a scale $M=1000\TeV$, each extra site
needs a coupling $\alpha_{sX}(M) \approx 0.26-0.35$, giving masses
$M_\Psi \sim 1 \TeV - 100 \TeV$, while for $N>4$ we find $M_\Psi \sim
M$.  The possible presence of these light vector-like colored fermions
with masses $M_\Psi \ll M$ generated by non-perturbative effects could
be an interesting signature of this theory to study in further work,
but for the remainder of this work we assume these states decouple and
focus on the details of the $SU(3)_1 \times SU(3)_2 \times SU(3)_3$
theory.

Another alternative possibility to avoid enlarging the gauge group
with extra $SU(3)$ factors is to consider a model with PQ symmetries
ensuring $y_d=0$ instead of $y_s=0$, so that smaller non-perturbative
effects are required to generate the quark mass ratios. This
possibility is appealing but is in tension with constraints on
$\bar\theta$, as described in~\cref{app:altyukawa}.

\subsection{The scalar sector}

The full description of the relevant particle content of the $SU(3)_1 \times SU(3)_2 \times SU(3)_3$ model is
given in \cref{tab:particles}.  There are several possibilities for the scalar
fields breaking the gauge group to the diagonal, here we take a simple
choice motivated by CKM mixings as described in the following
section.

\begin{table}
\centering
\begin{tabular}{l|rrr}
 & $SU(3)_i$ & $SU(3)_j$ & $(B-L)_i $ \\
 \hline
 $q_i$ 	 & $\square $ & 1 &$ \frac{1}{3}$ \\
 $u^c_i$ & $\bar\square $ &1& $ -\frac{1}{3}$\\
 $d^c_i$ & $\bar\square $ & 1&$-\frac{1}{3}$ \\
$\Sigma_{ij}$ 	& $\bar\square$	& $\square$& $\mp\frac13$ \\ 
 %\hline\rule{0pt}{4ex}
\end{tabular}
\caption{Particle content and charges under the family $SU(3)$
factors and family $B-L$ symmetries  for a
simple 3-generation model ($i,j=1,2,3$).}
\label{tab:particles}
\end{table}

We assume that the scalar link fields $\Sigma_{12}$, $\Sigma_{23}$,
and $\Sigma_{31}$ get vevs $f_{12} \sim f_{23}\sim f_{31}$ to break
the gauge group down to the diagonal, with $M$ corresponding to the
scale of Higgsing $M\sim g f$ (only two link fields are necessary to
break the gauge group, but the simplest renormalizable flavor models
will involve three link fields). The renormalizable potential allowed
by the symmetries  leads to spontaneous breaking of the gauge group
without introducing any new CP phases or uneaten light Goldstone boson
degrees of freedom. A standard renormalizable Higgs-like potential
drives a vev for each field, 
\begin{align}
V 
&= 
V_0(\Sigma_{12}) + V_0(\Sigma_{23}) + V_0(\Sigma_{31}) + 
  (\gamma {\rm Tr}(\Sigma_{12}\Sigma_{23}\Sigma_{31}) + h.c.), \\
 V_0(\Sigma_{ij}) 
  &= 
  %\lambda_\Sigma \left(|\Sigma_{a}|^2 - f^2\right)^2
 -m_{\Sigma_{ij}}^2 {\rm Tr}({\Sigma_{ij}\Sigma_{ij}^\dagger})
+\frac{\lambda_{ij}}{2}[{\rm Tr}({\Sigma_{ij}\Sigma_{ij}^\dagger})]^2
+\frac{\kappa_{ij}}{2}{\rm Tr}(\Sigma_{ij}\Sigma_{ij}^\dagger\Sigma_{ij}\Sigma_{ij}^\dagger) {\rm~(no~ sums~on~i,j)}
\label{eq:ScalarPotential}
\end{align}
The couplings $\lambda_{ij}$, $\delta_{ij}$, and $\lambda_{ij}$ are independent real parameters for each field $\Sigma$. The phase of $\gamma$ can be removed by a field redefinition, and causes the vacuum to align with vevs $\langle \Sigma_{12} \rangle$ , $\langle
\Sigma_{23} \rangle$ , $\langle \Sigma_{31} \rangle$ that can all consistently be chosen to be real.
Taking $\gamma$ to be a small perturbation for simplicity, we find \cite{Bai:2010dj,Bai:2017zhj}
\begin{equation}
\langle \Sigma_{ij} \rangle = \frac{m_{\Sigma_{ij}}}{\sqrt{\kappa_{ij}+3\lambda_{ij}}} \mathbb{I}_3 \equiv \frac{f_{\Sigma_{ij}}}{2} \mathbb{I}_3.
\end{equation} 
For simplicity assume all of the scales are comparable, $f_{12} \sim
f_{23} \sim f_{31} \sim f$, giving a common scale $M\sim g f$ cutting
off the instanton integrals in each $SU(3)$ factor.

 Renormalizable
cross-couplings of the form $\lambda' |\Sigma_{12}|^2|\Sigma_{23}|^2$,
$\lambda'' |\Sigma_{12}\Sigma_{23}|^2$, etc. are also allowed by the
symmetries, but do not  introduce any new CP phases and are
qualitatively unimportant as long they do not destabilize the vevs.
The symmetry breaking pattern at the level of the scalar potential so
far is $SU(3)^3 \times U(1)_{B-L}^3 \times U(1)_\Sigma \rightarrow
SU(3)\times U(1)_{B-L}$. The 16 colored Goldstone bosons are eaten by
the Higgs mechanism for the broken $SU(3)$ factors. The 8 remaining
colored pseudo-Goldstone bosons radiatively obtain masses at the scale
$\sim g^2 f^2$.   There remain three singlet Goldstones to lift. The
renormalizable term, $\gamma{\rm Tr}(\Sigma_{12}\Sigma_{23}\Sigma_{31}) + h.c.$
explicitly breaks the $U(1)_\Sigma$ symmetry, lifting one Goldstone.
Gauging two of the $U(1)_{B-L}$ factors can lift the remaining
Goldstones without introducing any new phases. Alternatively,
explicitly breaking the $U(1)_{B-L}$ factors with terms of the form
$\kappa' {\Sigma_{12}}^a_i{\Sigma_{12}}^b_j{\Sigma_{12}}^c_k
\epsilon_{abc}\epsilon^{ijk}$ would introduce new CP phases to the
theory, but in a controlled way for $\kappa'\ll f$. 

%The individual $SU(3)$ factors need to be approaching strong coupling
%at the scale of Higgsing so that the non-perturbative effects are
%large enough to generate the observed Yukawa couplings. Each factor
%may have a different value of $\alpha_{s_i}(M)$, accounting for the different
%observed up-down mass ratios for each generation.

\subsection{CKM and no $\bar\theta$ at tree level}
\label{sec:CKMhigherdim}

The model we have introduced so far generates the diagonal Yukawa
couplings and breaks the product gauge group down to the standard
model $SU(3)_c$, all while maintaining an accidental CP symmetry at
the renormalizable level. After integrating out the non-perturbative
effects near the scale $M$, the theory matches to the standard model
with non-vanishing diagonal Yukawa couplings for all of the quarks
and phases that preserve $\bar\theta=0$. 
\begin{align}
  D^u
  &= 
  \begin{pmatrix} 
    r_1 e^{i\theta} Y^{d*}_{11} &  & 0 \\
    0   & Y^u_{22} & 0 \\
    0   & 0 &  Y^u_{33}
  \end{pmatrix}
  \\
  D^d
  &= 
  \begin{pmatrix} 
    Y^d_{11} & 0 & 0\\
    0 &  r_2 e^{i\theta_2} Y^{u*}_{22} &  0 \\
    0   & 0 &  r_3 e^{i\theta_3} Y^{u*}_{33}
  \end{pmatrix}
  \label{eq:Yukdiag}
\end{align}
The PQ symmetries and large
non-perturbative effects are crucial to the accidental CP symmetry,
since they allow the breaking of the quark chiral symmetries without
introducing extra CP violating parameters. 

The next challenge is to introduce the CKM mixing without spoiling
this  protection.  When we introduce additional  off-diagonal Yukawa
couplings, the accidental CP symmetry can no longer survive, since the
observed CKM phase must be generated. However, $\bar\theta_{SM}$ will
still vanish at tree level and remain highly suppressed even at loop
level due to the residual approximate flavor symmetries.

Introducing quark-mixing between generations requires higher
dimensional operators involving the link fields, e.g.
\begin{align}
  \mathcal{L}_{d=5}
  &=
  \lambda^u_{ij} \left(q_i \frac{\Sigma_{ij}}{\Lambda_f} u^c_j\right)
  H \label{eq:yHDO}
\end{align}
generates the
effective Yukawa matrices
when the $\Sigma$ fields acquire vacuum expectation values. We can
write the off-diagonal Yukawa couplings below the scale of
Higgsing, 
\begin{align}
O^{u,d}
&=
\frac{\lambda^{u,d}_{ij} f_{ij}}{\Lambda_f}.
\label{eq:YOffDiagonal}
\end{align}
Since the off-diagonal entries in the Yukawa matrices can be small,
the flavor scale $\Lambda_f \gg M$  is possible, with a
separation as large as  $\Lambda_f \lesssim 10^4 M$
consistent with unitarity and the size of the observed off-diagonal
Yukawa elements. However, a natural assumption that the couplings
$\lambda$  of the UV completion are comparable to the non-vanishing
diagonal Yukawa couplings would require e.g. $\Lambda_f \sim f$ to
generate the $\mathcal{O}(1)$ Cabibbo angle.

\begin{table}
\centering
\begin{tabular}{l|rrr}
 &  $PQ_1$ & $ PQ_2$ & $PQ_3$  \\ \hline\hline
 %\rule{0pt}{4ex}
 $q_1$ 	& 		$0$ & $0$ & $-1$ \\
 $u^c_1$&	$1$ & $0$ & $0$ \\
 $d^c_1$&	$0$ & $0$ & $1$\\
 \hline
 %\rule{0pt}{4ex}
  $q_2$  & $0$ & $0$ & $-1$ \\
 $u^c_2$ & $0$ & $0$ & $1$ \\
 $d^c_2$ & $0$ & $1$ & $0$ \\
 \hline
 %\rule{0pt}{4ex}
  $q_3$  & $0$ & $0$ & $0$ \\
 $u^c_3$ & $0$ & $0$ & $0$ \\
 $d^c_3$ & $0$ & $0$ & $1$ 
 %\rule{0pt}{4ex}
\end{tabular}
\caption{Particle charges under family $PQ$ symmetries, for a
simple 3-generation model with CKM mixings ($i,j=1,2,3$).}
\label{tab:particlesCKM}
\end{table}

For general off-diagonal couplings, it is no longer true that the
tree-level $\bar{\theta}$ vanishes after matching to the SM,
\begin{align}
  \bar{\theta}
&=
\arg\det\left[
  e^{-i(\theta_1+\theta_2+\theta_3)}
  (D^u + O^u)
  (D^d + O^d)
\right]
\\&=
\arg
\left(
\det\left[
  e^{-i(\theta_1+\theta_2+\theta_3)}
  D^u D^d
\right]
\det \left[(1+  (D^{u})^{-1} O^u)\right]
\det \left[(1+  (D^{d})^{-1} O^d)\right]
\right)
%\\&=
%\arg
%\left(
%-\frac12 tr\left[(Y^{u})^{-1} y^u\right]^2
%+\det\left[(Y^{u})^{-1} y^u\right]
%-\frac12 tr \left[ (Y^{d})^{-1} y^d)\right]^2
%+\det \left[((Y^{d})^{-1} y^d)\right]
%\right)
\,.
\end{align}
The first determinant factor is real, as shown above. For the
other two factors,
it is
simple to see that we must require that the off-diagonal matrices 
$O^{u,d}$ can be put in a
strictly triangular form (up to $SU(3)$ rotations). We
would like the quarks to transform under
(possibly anomalous) $U(1)$ symmetries that perturbatively protect
this form, and in fact there are only two possible textures satisfying
these constraints and giving viable CKM mixings. The texture we will
focus on is:
\begin{align}
  Y^u  
  &= 
  \begin{pmatrix} 
    0   & Y^u_{12} & 0 \\
    0   & Y^u_{22} & 0 \\
    0   & 0 &  Y^u_{33}
  \end{pmatrix}
  \\
  Y^d  
  &= 
  \begin{pmatrix} 
    Y^d_{11} & 0 & Y^d_{13}\\
    Y^d_{21} & 0 & Y^d_{23} \\
    0   & 0 &  0
  \end{pmatrix},
  \label{eq:UVYuks}
\end{align}
and the assignment of PQ charges in~\cref{tab:particlesCKM} protects this
form of the Yukawa matrix. The other possible texture, described briefly in~\cref{app:altyukawa}, gives a less natural realization of the CKM structure.

The three anomalous $U(1)_{PQ}$ symmetries allow us to rotate away the
$\theta$ angle in each $SU(3)$ factor, and field redefinitions leave
only two remaining physical CP phases in the Yukawa matrix, which we
choose by convention to put in the $Y^d_{23}$ and $Y^d_{21}$ elements.
Including non-perturbative instanton effects and for the moment
ignoring all other radiative effects, below the scale $M$ the theory
matches to the SM with Yukawa matrices

\begin{align}
y^u  
&= \begin{pmatrix} 
r_1 e^{i\theta_1} {Y^d_{11}}^*    & Y^u_{12} & 0 \\
0   & Y^u_{22} & 0 \\
0   & 0 &  Y^u_{33}
\end{pmatrix}
\\
y^d  &= \begin{pmatrix} 
 Y^d_{11} & 0 & Y^d_{13}\\
 Y^d_{21} & r_2 e^{i\theta_2} {Y^u_{22}}^* & Y^d_{23} \\
0   & 0 &  r_3 e^{i\theta_3} {Y^u_{33}}^*
\end{pmatrix}
\end{align}
We can check explicitly that $\bar\theta_{SM}=0$ at tree level,
\begin{align}
  \bar{\theta}
&=
-\arg\det\left[
  e^{-i(\theta_1+\theta_2+\theta_3)}
\begin{pmatrix} 
r_1 e^{i\theta_1} {Y^d_{11}}^*    & Y^u_{12} & 0 \\
0   & Y^u_{22} & 0 \\
0   & 0 &  Y^u_{33}
\end{pmatrix}
\begin{pmatrix} 
 Y^d_{11} & 0 & Y^d_{13}\\
 Y^d_{21} & r_2 e^{i\theta_2} {Y^u_{22}}^* & Y^d_{23} \\
0   & 0 &  r_3 e^{i\theta_3} {Y^u_{33}}^*
\end{pmatrix}
\right]
=
0
\,.
\end{align}
The real
coefficients $r_{1,2,3}$ parameterize the size of the instanton
suppression of PQ breaking in each $SU(3)$ factor. The couplings $Y$ can
now be determined from the CKM matrix and the observed SM fermion
masses. The only undetermined parameter is $Y^d_{21}/Y^d_{11}$, but we
will be motivated shortly to focus on the limit $Y^d_{21}\ll
Y^d_{11}$. Then to leading order in the small Yukawa ratios
$y_{u,d}/y_{c,s,t,b}$, $y_{s}/y_{b}$, and small CKM mixings
$|V_{31}|= 0.0089, |V_{32}|=0.041$ \cite{Agashe:2014kda}  we obtain
\begin{align}
Y^u_{33} &=y_t, Y^u_{22}= y_c \cos\theta_c, Y^u_{12}=y_c \sin\theta_c
\nonumber \\
Y^d_{11}&=y_d, Y^d_{13}= y_b |V_{31}| e^{i\delta_0}, Y^d_{23} = y_b |V_{32}|
\nonumber\\
r_1 &= \frac{y_u}{\cos\theta_c y_d}, r_2 = \frac{y_s}{\cos\theta_c
y_c}, r_3 = \frac{y_b}{y_t}
\,,
\end{align}
where we have made a field definition choice to put the CKM phase
entirely into $Y^d_{13}$ and $\theta_c$ is the Cabibbo angle. An
alternative solution with the same texture but flipping the role of
the strange and down quarks is discussed in \cref{app:altyukawa}.

Now that the CKM elements are introduced, the gauge basis in the
$SU(3)\times SU(3)\times SU(3)$ theory is no longer aligned with the
flavor basis, and four-fermion operators generated by gauge
interactions at the scale $M$ will introduces non-MFV contributions to
CP-preserving flavor observables. The dominant constraint is due to
the $\Delta C=2$
operator generated by exchange of the heavy broken SU(3) gauge bosons,
given in the quark mass basis as 
\begin{align}
  \mathcal{O}_{\Delta C=2}
  \sim\frac{4\pi \alpha_{s_{1,2}}
\sin^2\theta_C}{M^2} (\bar c \gamma^\mu u)^2.
\end{align}
Constraints on the $D^0$ splitting generated by this operator give
$M\gtrsim 1000\TeV$~\cite{Isidori:2010kg}. The
leading $\Delta B = 2$ and $\Delta S =2$ operators are suppressed
respectively by $|V_{13,23}|^2$  and $|V_{12,23}y_d / y_s|^2$  and
give less stringent constraints.

\subsection{$\Delta\bar{\Theta}$ from thresholds }

With the two physical CP violating phases in the $Y^d_{23}$ and
$Y^d_{12}$ elements, it is clear that at leading order in the Yukawa
couplings, neither contributes to the low energy theta angle. However
higher order perturbative corrections to the non-perturbative effects
at $M$ can give a non-vanishing threshold correction to
$\bar\theta_{SM}$.

At energies below $M$, the additional breaking of the SM
flavor symmetry generated by the gauging of $SU(3)_1 \times SU(3)_2
\times SU(3)_3$  decouples and the theory is just the standard model,
where the flavor symmetries suppress the running of  $\bar\theta$ to
negligible effects starting at 7-loops
\cite{Ellis:1978hq,Dugan:1984qf}. At energies far above
$M$, the non-perturbative PQ violating effects are
exponentially suppressed by the weak coupling of the gauge groups, and
the PQ symmetry protects the form of the Yukawa matrices with
$\bar\theta=0$ manifest, \cref{eq:UVYuks}. Therefore the
dominant effect on $\bar\theta$ is a threshold effect at energies near
$M$, where the non-perturbative violation of the PQ
symmetries are still large and the extra breaking of the SM flavor
symmetries through the gauging of $SU(3)_1 \times SU(3)_2 \times
SU(3)_3$ has not decoupled. 

\begin{figure}
  \centering
\includegraphics[width=4in]{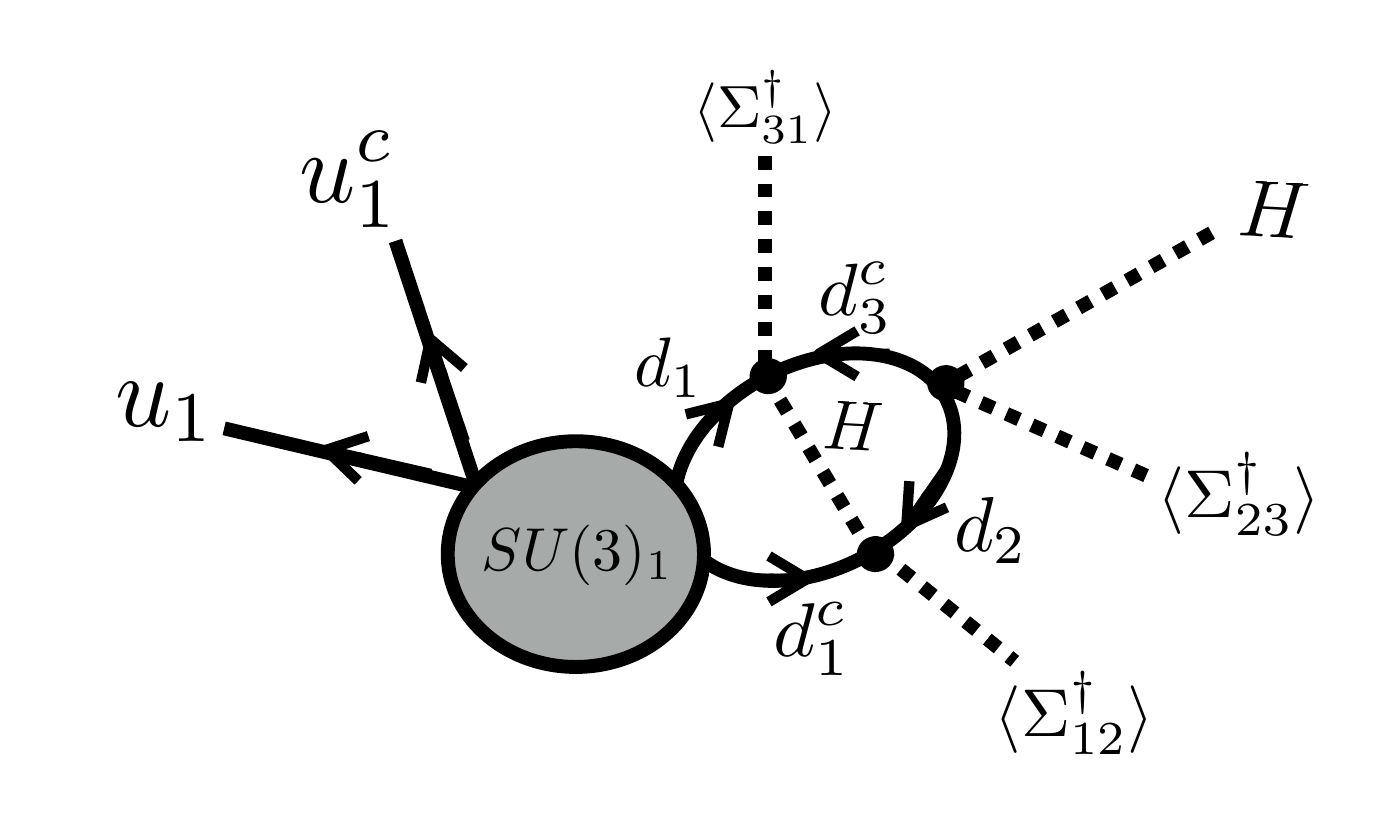}
\caption{One of the leading diagrams generating a non-vanishing
  threshold correction to $\Delta\bar\theta$. The off-diagonal Yukawa
  couplings appear in order to introduce a CP phase, and the instanton
  violates the anomalous PQ symmetry protecting the UV form of the
  Yukawa
  couplings as in~\cref{eq:UVYuks}.}
\label{fig:ThetaThreshold}
\end{figure}

The leading effects occur at third order in the Yukawa couplings,
schematically generated from diagrams of the form
of~\cref{fig:ThetaThreshold}.
Roughly, these diagrams describe how the Yukawa elements closing the
instanton diagrams depend on the scale of the instanton -- there is a
mismatch of the phase between instantons at different scales because
of the perturbative running of the Yukawas. Taking the $\Sigma$ fields
as background fields, the 1-loop running of the effective Yukawa
couplings \cref{eq:YOffDiagonal} takes the same form as in the SM
\cite{Machacek:1983fi}, with the non-vanishing CP phases entering
through the terms
3rd order in the Yukawa couplings:

\begin{align}
\beta_{P}(Y_{u,d}) 
&= \frac{1}{16\pi^2}\frac{3}{2}(Y_{u,d} 
Y_{u,d}^\dagger - Y_{d,u}Y_{d,u}^\dagger)Y_{u,d}
\label{eq:PertRunning}
\end{align}
Since the phases entering in the instantons no longer align exactly with the low energy perturbative values of the Yukawa couplings, there is no longer an exact cancellation in phase between the non-perturbatively generated eigenvalues and the perturbative eigenvalues of $Y_{u,d}$. 

To obtain a parametric estimate of these effects, we iteratively solve
the RGE including the perturbative running~\cref{eq:PertRunning} and
non-perturbative running~\cref{eq:yrun} of the Yukawas, as described
in detail in \cref{app:yukawarge}. We ignore the effects of
perturbative gauge interactions and the propagation of the $\Sigma$
fields -- all effects that modify $\bar\theta$ must involve both an
instanton and a Yukawa loop, so these higher order effects can give at
most $\mathcal{O}(1)$ corrections to our estimate if these states are strongly
coupled. Finite effects not captured by the RGE are also expected to
be of comparable size. There are two leading contributions to
$\bar\theta$. The size of the first is fixed by the experimentally
determined elements of the Yukawa matrix, 
\begin{align}
\Delta\bar\theta 
&= \frac{3}{16 \pi^2} \frac{1}{b} 
\tan\theta_C V_{31} V_{32}
y_b^2\sin(\delta_0)[f_I(\alpha_{s_1}) + f_I(\alpha_{s_2})]  
\label{eq:deltatheta1}
\\
&\approx  10^{-10} \times 
\left(\frac{f_I(\alpha_{s_1}) +
f_I(\alpha_{s_2})}{2}\right)
\end{align}
where 
\begin{align}
f_I(\alpha)
&= \frac{\Gamma(8, 2\pi\alpha^{-1})}
{\Gamma(7, 2\pi\alpha^{-1})}-2\pi\alpha^{-1}
\approx 0.8 + 1.6\alpha
\label{eq:fI}
\end{align}
with this linear approximation holding well in the coupling region of
interest $\alpha \sim 0.2-1$.
The small size of $\Delta\bar\theta \sim 10^{-10}$ is due to the loop
suppression and the smallness of the off-diagonal Yukawa elements. The
form of $\Delta\bar\theta$ is consistent with the observation that
$Y_{13}$ and $Y_{23}$ must appear as a product, since the  physical
phase can be rotated from one term to the other. The suppression by a
factor of $1/b = 3/29$ arises because there is only a small range of
energies where instanton effects are important, controlled by how
rapidly the gauge coupling runs. 

There is another contribution proportional to the undetermined Yukawa
element $Y_{21}$,
\begin{align}
\Delta \bar \theta' 
&= -\frac{3}{16\pi^2}\frac{1}{b} 
\left( \cos\theta_C \sin\theta_C y_c^2 {\rm Im} [Y^d_{21}]
[f_I(\alpha_{s_1}) + f_I(\alpha_{s_2})]
\right)
\nonumber \\&
\approx 
-(4\times10^{-8}) \times \frac{{\rm Im}[Y^d_{21}]}{y_d}
\left(\frac{f_I(\alpha_{s_1}) + f_I(\alpha_{s_2})}{2}\right)
\label{eq:deltatheta2}
\end{align}
If $Y^{d}_{21}$ takes on a value $\sim y_d$  with $\mathcal{O}(1)$ phase, this
extra contribution is inconsistent with experimental limits. However,
spurion arguments show that  $|Y^d_{21}|\ll y_d$ can be naturally
obtained. Since $Y^d_{21}$ breaks a different set of flavor
symmetries, its natural size can  be as small as 
\begin{align}
|Y^d_{21}|\gtrsim Y^d_{11} 
\times {\rm Max}({Y^d_{13}}^* Y^d_{23}, {Y^u_{12}}^* Y^u_{22}) 
\approx y_d  
\times {\rm Max}(y_b^2 V_{31} V_{32}, y_c^2 \sin\theta_C \cos\theta_C) 
\approx y_d\times 10^{-5} 
\end{align}
making $\Delta\bar\theta'$ subdominant. 

We have checked these estimates numerically at the one-loop level.

\subsection{UV Sensitivity}
\label{sec:UVsensitivity}

It is useful to discuss the degree to which this mechanism is
insensitive to ultraviolet physics at some scale $\Lambda_{UV}$ where
CP may be violated in a sector strongly coupled to the standard model.
For CP violation to be communicated from this sector to $\bar\theta$,
the anomalous breaking of the PQ symmetry must be active. There are two
possible contributions: small instantons of scale $\Lambda_{UV}^{-1}$
interacting directly with the new UV physics, and the unsuppressed
instantons at the scale $M^{-1}$ interacting with the physics at
$\Lambda_{UV}$ through higher dimensional operators.

The contributions of small instantons of size $\Lambda_{UV}^{-1}$ is
suppressed by the exponentially small instanton density
$D(\Lambda_{UV}^{-1})$ as long as the individual $SU(3)$ factors have
run back to weak coupling. For example, suppose the physics at
$\Lambda_{UV}$ introduces an $\mathcal{O}(1)$ phase $\alpha$ in the
non-vanishing Yukawas, e.g. $y_d(\Lambda_{UV}) \approx e^{i\alpha}
y_d(M)$. Then instantons at the scale generate a
contribution to $y_u$ with a phase that will appear in $\bar\theta$,
\begin{align}
\frac{{\Delta y_u}_{UV}} {y_u} 
&\sim 
e^{i\alpha} \frac{D(\Lambda_{UV}^{-1})}{D(M^{-1})}
\end{align}
In a sector with two-flavors, this contribution is consistent with
$\Delta\bar\theta \lesssim 10^{-10}$ if $\Lambda_{UV} \gtrsim 100 M$.

The physics at the scale $\Lambda_{UV}$ can also generate higher
dimensional operators consistent with the PQ symmetries and other
approximate chiral symmetries that carry CP phases and can interact
with the unsuppressed instantons at the scale $M$ (such operators also
interact with instantons at the scale $\Lambda_{QCD}$ and generate a
shift in $\bar\theta$ even in the standard PQ axion or massless up quark solution
\cite{Hamzaoui:1998yu}, but here these effects are subdominant by a
factor $\Lambda_{QCD}^2/M^2$). The most dangerous
operators are momentum dependent contributions to the phase of the
perturbatively allowed diagonal Yukawas,
\begin{align}
  \mathcal{L}_{d=6}  
  &\supset 
  {Y_{11}^d}' H^\dagger Q_1 \frac{D_\mu^2}{\Lambda_{UV}^2} d^c_1
  +  {Y_{22}^u}' H Q_2 \frac{D_\mu^2}{\Lambda_{UV}^2} u^c_2
  + {Y_{33}^u}' H Q_3 \frac{D_\mu^2}{\Lambda_{UV}^2} u^c_3
  \label{eq:D6}
\end{align}
Combined with instanton insertions at $M$, these give
\begin{align}
\Delta\bar\theta 
&\sim 
{\rm Im}\left[\frac{Y'}{Y}\right]\frac{{M}^2}{\Lambda_{UV}^2}
\,.
\end{align}
When $Y' \sim Y$ and
the phases are uncorrelated, this requires $\Lambda_{UV} \gtrsim 10^5 M$
to avoid generating $\Delta \bar\theta$. 

Another dangerous $d=6$ operator that can generate contributions to
$\bar\theta$ even in the absence of PQ breaking are mixed topological
terms, for example
\begin{align}
\mathcal{L}_{d=6} 
&\supset \hat\theta_{12} G^{(1)\;b}_{\;a}\tilde{G}^{(2)\; i}_{j} \frac{ {\Sigma_{12}}_b^{\;j} {\Sigma_{12}^\dagger}^a_{\;i}}{\Lambda_{UV}^2}
\label{eq:D6mixedtopo}
\end{align}
gives a contribution $\Delta\bar\theta \approx \hat{\theta}_{12}
\frac{M^2}{\Lambda^2_{UV}}$, again requiring
$\Lambda_{UV} \gtrsim 10^5 M$ unless $\hat\theta_{12}$ is suppressed.

\section{A Flavor UV Completion}
\label{sec:UVFlavor}

The $d=5$ operators in \cref{eq:yHDO} generating the off-diagonal
Yukawa elements require a UV completion at the scale $\Lambda_f$.
Unitarity of the $d=5$
operator in \cref{eq:yHDO} generating the off-diagonal Yukawas
requires $\Lambda_f \lesssim 10^{-4} M$. Taking the effective action to $d=6$ introduces operators consistent
with the PQ symmetries that could allow the CP violation generating
$\delta_{CKM}$ to enter directly into $\Delta\bar\theta$, as
discussed in~\cref{sec:UVsensitivity}.  

In this section we give an example of a simple UV completion in which
the higher dimension operators do not make large contributions to
$\Delta
\bar\theta$  and which can also explain the origin of the spurion
argument giving $|Y^d_{21}|\ll y_d$ . The model is extended to involve
a set of vector-like fermions $Q_3, \bar{Q}_3, U^c_1, \bar{U^c_1}$,
with charges under the gauge and PQ symmetries as given
in~\cref{tab:vectorlikeCKM}. Renormalizable mixings between heavy
states and the SM-like fields generates the higher dimensional
operators~\cref{eq:yHDO} after integrating out the vector-like
states.

\begin{table}
\centering
\begin{tabular}{l|lll}
 & \tiny ${SU(3)_1}_{(B-L)_1, PQ_1}$ &\tiny ${SU(3)_2}_{(B-L)_2, PQ_2}$ &\tiny ${SU(3)_3}_{(B-L)_3, PQ_3}$  
 \\
 \hline\hline\rule{0pt}{4ex}
 $U^c_1$		& $\bar\square_{(-\frac{1}{3}, 0)}$ & - & $-_{(0,1)}$ 
 \\
 $\bar{U^c_1}$		& $\bar\square_{(-\frac{1}{3}, 0)}$ & - & $-_{(0,-1)}$   \\ 
 \hline\rule{0pt}{4ex}
  $Q_3$ 			& - & - & $\square_{(\frac{1}{3},-1)}$   \\
  $\bar{Q_3}$ 			& - & - & $\square_{(\frac{1}{3},1)}$   \\ 
\end{tabular}
\caption{Vector-like quark content allowing a simple UV completion of
the off-diagonal higher dimension CKM mixing operators.}
\label{tab:vectorlikeCKM}
\end{table}

The renormalizable terms in the Lagrangian consistent with the gauge
and PQ symmetries are
\begin{align} 
  \mathcal{L}_{\rm UV} 
  &= 
  M_U U_1^c
  \bar{U^c_1} + M_Q Q_3 \bar{Q_3} 
  \nonumber\\& \qquad
  + z^u_{11} H q_1 U_1^c  
  + z^d_{33} H^\dagger Q_3 d_3^c 
  \nonumber\\ & \qquad
  +x^U_{12} \Sigma_{12} u^c_2 \bar{U^c_1} 
  +x^Q_{13} \Sigma_{13} q_1 \bar{Q_3} 
  +  x^Q_{23} \Sigma_{23} q_2 \bar{Q_3} 
  \nonumber\\& \qquad
  +{\hat{Y}}^d_{11} H^\dagger q_1 d_1^c 
  +{\hat{Y}}^u_{22} H q_2 u_2^c
  + {\hat{Y}}^u_{33} H q_3 u_3^c
\end{align}
The field redefinition freedom leaves one physics CP violating phase
in this Lagrangian, which can be rotated between the parameters $M_U,
x^u_{12}, x^Q_{13}, x^Q_{23}, z^u_{11}, \hat{Y}^u_{22}$.  Taking
$M_{Q,U}\gg M $ and integrating out these states at tree level, we
obtain the effective theory of~\cref{sec:CKMhigherdim}, with
couplings to leading order in $\langle \Sigma \rangle / M$
\begin{align}
& Y^d_{11}={\hat Y}^d_{11}, \;\;Y^u_{22}={\hat Y}^u_{22}, \;\; Y^u_{33}={\hat Y}^u_{33} \\
& Y^u_{12}=\frac{x^U_{12}\langle \Sigma_{12} \rangle}{M_U} z^u_{11} \\
& Y^d_{13}=\frac{x^Q_{13}\langle \Sigma_{13} \rangle}{M_Q} z^d_{33}, \;\; Y^d_{23}=\frac{x^Q_{23}\langle \Sigma_{23} \rangle}{M_Q} z^d_{33}, \\
& Y^d_{21}= {\hat Y}^d_{11}  \frac{x^Q_{13}\langle\Sigma_{13}\rangle (x^Q_{23} \langle\Sigma_{23}\rangle)^\dagger}{|M_Q|^2}
\end{align}
By convention we can rotate the physical phase entirely into
$Y^d_{23}$ and $Y^d_{21}$ in the low energy Yukawa matrix, and to
leading order this corresponds to rotating the phase entirely into
$x_{23}^Q$ in the full theory. 

The contribution to $\Delta\bar{\theta}$ due to $Y_{21}$
(\cref{eq:deltatheta2}) is suppressed by the mixing of $q_1$ and
$q_2$ with the vectorlike $Q_3$,
\begin{align}
\left|\frac{x^Q_{13}\langle\Sigma_{13}\rangle (x^Q_{23}
\langle\Sigma_{23}\rangle)^\dagger}{|M_Q|^2}\right|
&=
\left|\frac{Y^d_{13}{Y^d_{23}}}{{z^d_{33}}^2}^\dagger\right| 
\approx 
3 \times 10^{-7} \times \left(\frac{1}{|z^d_{33}|^2}\right),
\end{align}
giving
\begin{align}
\Delta\bar\theta'
&\approx 
10^{-15} \times \left(\frac{1}{|z^d_{33}|^2}\right)
\end{align}
which requires the coupling $|z^d_{33}|\gtrsim 10^{-2}$ for this to be
subdominant. 

A more dangerous contribution in this model comes from the $d=6$
operators of \cref{eq:D6} (the contributions from mixed topological
terms~\cref{eq:D6mixedtopo} are subdominant). There are
unsuppressed terms generated a tree level giving $Y'^u_{22} \sim
Y^u_{22}$ and $Y'^d_{11} \sim Y^d_{11}$, with scale $\Lambda_{UV} =
M_Q$. However, these operators do not contribute to $\Delta\bar\theta$
because the phases of $Y$ and this contribution to $Y'$ are aligned.
The leading contribution with a misaligned phase is
\begin{align}
{\Delta Y^u_{22}}' &= Y^u_{12}
\frac{x^Q_{13}\langle\Sigma_{13}\rangle (x^Q_{23}
\langle\Sigma_{23}\rangle)^\dagger}{|M_Q|^2}=Y^u_{12}\frac{Y^d_{13}{Y^d_{23}}^\dagger}{|z^d_{33}|^2}
\end{align}
This term is also suppressed by the mixing of $q_1$ and $q_2$ with
the vectorlike $Q_3$. For
$\langle\Sigma_{13}\rangle\sim\langle\Sigma_{23}\rangle \sim M$, the
contribution to the theta angle is 
\begin{align}
\Delta\bar{\theta}_{D=6}\sim
\frac{{{\rm Im}[\Delta Y^u_{22}}']}{Y^u_{22}} \frac{{M}^2}{|M_Q|^2}
\sim \frac{Y^u_{12}}{Y^u_{22}}
|Y^d_{13}|^2|Y^d_{23}|^2\frac{g^2}{x^Q_{13}x^Q_{23}} \sin\delta_0
\approx
10^{-13}\times\left(\frac{\alpha_{s_{1,2,3}}}{0.5}\right)\left(\frac{1}{|x^Q_{13}x^Q_{23}
{z^d_{33}}^2| }\right) 
\end{align}
As long as the marginal couplings
generating the $q_1$ and $q_2$  mixings are not too weakly coupled,
$x^Q_{13}, x^Q_{23}, z_{33} \gtrsim 0.2$, this contribution is
subdominant. This corresponds to a rough lower limit on the scale
$M_Q \gtrsim 100 M$.  Note that a hierarchy $M_Q \gg M_U \sim M$ can naturally explain the small third-generation quark mixings and $O(1)$ Cabibbo angle.

This flavor model has a similar structure to minimal Nelson-Barr
models \cite{Barr:1979as,Nelson:1983zb,Barr:1984qx}, which obtain CKM
mixings through vector-like quarks \cite{BENTO199195}, forbidding a
tree-level $\bar \theta$. However, in contrast to the present case
where the $U(1)$ symmetries are sufficient to protect the structure of
the theory, in  Nelson-Barr models discrete symmetries and additional
UV structure are required \cite{Vecchi:2014hpa, Dine:2015jga}. In both
cases, radiative contributions to $\bar\theta$ limit the allowed
parameter space, but in Nelson-Barr models these limits appear to
generically require unexplained suppressions of allowed couplings
\cite{Dine:2015jga}.

\section{Conclusions}

The solutions to the strong CP problem and the origins of the flavor
structure of the standard model may be intricately tied to each other.
In this work, we constructed a model where embedding QCD in a
$SU(3)^3$ gauge group with flavorful anomalous PQ symmetries can
naturally explain the non-observation of a neutron EDM, the smallness
of the third generation CKM mixing angles, and the relative
suppression of the down-like quark masses in the second and third
generation. 
The theta angle in each $SU(3)$ factor can be set to zero using an
anomalous PQ symmetry. This symmetry is realized by forbidding a bare
mass for the lighter quark in each generation (i.e.\,$u,s,b$). Their
masses are generated through instantons, dominantly at the scale of
$SU(3)^3$ breaking, $M$, which can be far above the weak scale. The
instanton-generated mass terms have phases
that are naturally aligned with the theta angle, and hence do not
reintroduce a non-zero $\bar{\theta}$. 

There is a non-zero $\bar{\theta}$ generated at the threshold $M$ through
loop corrections that involve both the instanton vertex as well as the
perturbative CKM phase.
In fact, in our model the smallness of the CKM mixing angles is
intimately tied to the smallness of $\bar\theta$ in this model, and
the observed CKM elements give a prediction $\bar\theta \sim 10^{-10}$
that can be probed at the next generation of neutron EDM
\cite{Ito:2007xd,Tsentalovich:2014mfa} and proton storage ring
experiments \cite{Anastassopoulos:2015ura}.  The solution to the
strong CP problem is in the spirit of the massless up quark solution,
and there are no axion-like states in the theory.
An interesting future direction would be to study models which
generate the full standard model flavor structure
while also implementing our mechanism to solve the strong CP problem.

We would like to thank Asimina Arvanitaki, Savas Dimopoulos, Bogdan
Dobrescu, Howard Georgi, Roni Harnik, Anson Hook, Junwu (Curly) Huang,
Gustavo Marques Tavares, Lisa Randall, Matt Reece and Raman Sundrum
for encouragement and helpful comments. This work was initiated at the
Aspen Center for Physics, which is supported by National Science
Foundation grant PHY-1066293.  PA is supported by the NSF grants
PHY-0855591 and PHY-1216270.  This manuscript has been authored by
Fermi Research Alliance, LLC under Contract No.  DE-AC02-07CH11359
with the U.S. Department of Energy, Office of Science, Office of High
Energy Physics. The United States Government retains and the
publisher, by accepting the article for publication, acknowledges that
the United States Government retains a non-exclusive, paid-up,
irrevocable, world-wide license to publish or reproduce the published
form of this manuscript, or allow others to do so, for United States
Government purposes.

\appendix
\section{$\bar\theta$  Threshold}
\label{app:yukawarge}

The RGE can be written as an integral equation and solved iteratively
to obtain the threshold corrections to $\bar\theta$,
\begin{align}
  Y^{u,d}_{ij}(t) 
  &= 
  Y^{u,d}_{ij}(t_0) 
  +\int_{t_0}^t dt' 
  \left[ \beta_P[Y^{u,d}_{ij}](Y^u(t'),Y^d(t')) 
  + \beta_I[Y^{u,d}_{ij}](Y^u(t),Y^d(t), t') 
\right]
\end{align}
where the perturbative beta function depends only on $t'$ through the running of the couplings
\begin{align}
\beta_P[Y^{u,d}_{ij}](Y^u(t'),Y^d(t')) = 
\frac{1}{16\pi^2} \frac{3}{2} \left( Y^{u,d}(t')  {Y^{u,d}(t')}^\dagger  -  Y^{d,u}(t')  {Y^{d,u}(t')}^\dagger \right)_{i}^{~k}  Y^{u,d}(t')_{k j}
\end{align}
and the instanton contribution to the running depends explicitly on
the scale through the instanton density
\begin{align}
\beta_I[Y^{u,d}_{i=j}](Y^u(t'),Y^d(t'), t') 
&= 
-c_0 D[t'] {Y^{d,u}_{i=j}(t')}^*
\nonumber\\
\beta_I[Y^{u,d}_{i\neq j}] &= 0 
\end{align}
The leading terms in the series are given by
\begin{align}
Y^{u,d}(t) &= Y^{u,d}(t_0) + Y^{u,d~(I)}(t) +  Y^{u,d~(P)}(t) + Y^{u,d~(IP)}(t) +  Y^{u,d~(PI)}(t) + ...
\label{eq:Yseries}
\end{align}
where 
\begin{align}
  Y^{u,d~(P)}(t) 
  &= \int_{t_0}^t dt' \beta_P[Y^{u,d}_{ij}](Y^u(t_0),Y^d(t_0)) 
  \nonumber\\&=
  \frac{2}{b}\left(2\pi\alpha^{-1}(t)- 2\pi\alpha^{-1}(t_0)\right)\beta_P[Y^{u,d}_{ij}](Y^u(t_0),Y^d(t_0))
  \nonumber\\
  Y^{u,d~(I)}(t) &= \int_{t_0}^t dt' \beta_I[Y^{u,d}_{ij}](Y^u(t_0),Y^d(t_0), t') 
  \nonumber\\&
  = -\frac{2 c_0 D_0}{b} \Gamma\left(7, 2\pi\alpha^{-1}(t), 2\pi\alpha^{-1}(t_0)\right)
\end{align}
are the linear terms in the series. Note that although the dependence
of the Yukawa couplings on scale has been discarded under the integral, the
explicit $t$ dependence of the instanton contribution is maintained.
The leading cross terms between the instanton and perturbative running
enter at second order,
\begin{align}
  Y^{u,d~(PI)}_{ij}(t) 
  &= 
  - Y^{u,d~(P)}_{ij}(t)
  +
  %\left[
    \int_{t_0}^t dt' \beta_P[Y^{u,d}_{ij}](Y^{u}(t_0) 
    + Y^{u~(I)}(t'),Y^{d}(t_0) 
    +Y^{d~(I)}(t'))
  %\right] 
  \nonumber \\
  Y^{u,d~(IP)}_{i=j}(t) 
  &= 
  - Y^{u,d~(I)}_{i=j}(t)
  +
  %\left[
    \int_{t_0}^t dt' \beta_I[Y^{u,d}_{i=j}](Y^{u}(t_0) 
    + Y^{u~(P)}(t'),Y^{d}(t_0) 
    +Y^{d~(P)}(t'), t')
  %\right] 
  \nonumber \\
  Y^{u,d~(IP)}_{i\neq j}(t) 
  &= 0
\end{align}
where only terms linear in $Y^{I}, Y^{P}$ are to be kept.
$Y^{u,d~(IP)}(t)$ takes a simple form and is illustrative to examine.
In the limit $\alpha(t_0) \ll 1$,
\begin{align}
Y^{u,d~(IP)}_{i=j}(t) 
&= Y^{u,d~(I)}_{i=j}(t) \frac{Y^{d,u~(P)}_{i=j}(t)^*}{Y^{d,u}_{i=j}(t_0)^*}
\left(1 - \frac{f_I(\alpha(t))}{2\pi \alpha^{-1}(t_0) -
2\pi\alpha^{-1}(t)} \right)
\end{align}
where $f_I(\alpha(t))$ is given in~\cref{eq:fI}.  The term constant
in the parentheses has a simple interpretation; we can write
\begin{align} 
  Y^{u,d~(I)}_{i=j}(t) + Y^{u,d~(IP)}_{i=j}(t) =
  Y^{u,d~(I)}_{i=j}(t) \frac{Y^{d,u}_{i=j}(t_0)^* +
  Y^{d,u~(P)}_{i=j}(t)^*}{Y^{d,u}_{i=j}(t_0)^*} + \ldots 
\end{align}
Clearly the physical effect of this term is to shift the Yukawa
coupling entering the instanton to its value at the IR scale $t$
where instanton effects are large, not $t_0$ where the theory is
weakly coupled. This sets $\Delta\bar\theta=0$. 
The second term in the parentheses is a threshold effect -- since
\begin{align}
  Y^{d,u~(P)} 
  &\propto 
  (2\pi \alpha^{-1}(t_0) - 2\pi\alpha^{-1}(t))
  \,,
\end{align}
it
is finite while $2\pi\alpha^{-1}(t_0) \rightarrow \infty$  as $t_0
\rightarrow \infty$.  This captures the fact that the instantons are
active over a finite range of scale near the IR scale $t$, and are
sensitive to changes in the phases of the Yukawas near $t$. This
spoils the exact cancellation that set $\Delta \bar \theta  =0$ at
leading order. The other term $Y^{u,d~(PI)}$ describes similar
threshold effects as the instanton-generated Yukawas themselves are
rotated by the perturbative running.
Evaluating $\Delta\bar\theta$
with these leading terms in the series~\cref{eq:Yseries} gives the
results~\cref{eq:deltatheta1,eq:deltatheta2}.

\section{Alternative Yukawa Structures}
\label{app:altyukawa}

An alternative solution to the observed quark masses and CKM matrix is
possible with the Yukawa texture of \cref{eq:UVYuks} by switching the
role of $y_d$ and $y_s$. In this case the non-perturbative effects
generate $y_d$ from $y_c$, $y_u$ from $y_s$, and $y_b$ from $y_t$.
This is an attractive possibility because it requires smaller
non-perturbative effects, and therefore can more easily be accomodated
without adding additional weakly coupled sites to the $SU(3)\times
SU(3) \times SU(3)$ model. However, the size of the radiative
contribution to $\Delta \bar\theta$ is increased by a factor of
$(\cot\theta_c)^2 \approx 20$ in this model, which is excluded by
current limits unless there is a $\sim 10\%$ tuned cancellation with
another contribution to $\bar\theta$.

While we focused on the Yukawa texture \cref{eq:UVYuks}, there is one other possibility for a viable Yukawa texture that can be protected by $U(1)_{PQ}$ symmetries and has a vanishing tree-level contribution to $\bar\theta$,
\begin{align}
  Y^u  
  &= 
  \begin{pmatrix} 
    0   & 0 &  Y^u_{13}  \\
    0   & Y^u_{22} & 0 \\
    0   & 0 &  Y^u_{33}
  \end{pmatrix}
  \\
  Y^d  
  &= 
  \begin{pmatrix} 
    Y^d_{11} & Y^d_{12} & 0\\
    0 & 0 & 0   \\
     Y^d_{31}   & Y^d_{32} &  0
  \end{pmatrix},
  \label{eq:UVYuks2}
\end{align}
The CKM structure emerges less naturally for this texture  because of the right-handed dominant mixing structure in the down Yukawa matrix. Fitting the $V_{31}$ and $V_{32}$ CKM elements requires a cancellation between terms of order $Y^u_{31}/y_t$ and $Y^d_{12} Y^d_{32}/Y_b^2$. Nonetheless this texture remains an interesting possibility, and viable models can be realized and also generically predict $\bar\theta \sim 10^{-10}$ from the radiative corrections.

\bibliographystyle{utphys}
\bibliography{FlavorfulInstantons.bib}
\end{document}